\documentclass[aps,prb,amsmath,amssymb,reprint,superscriptaddress]{revtex4-1}

\usepackage{graphicx}
\usepackage{epstopdf}
\usepackage{natbib}

\begin{document}


\title{An Exact Diagonalization Study of the Anisotropic Triangular Lattice Heisenberg Model Using Twisted Boundary Conditions}


\author{Mischa Thesberg}
\email[]{thesbeme@mcmaster.ca}
\author{Erik S. S{\o}rensen}
\email[]{sorensen@mcmaster.ca}
\homepage[]{http://comp-phys.mcmaster.ca}
 \affiliation{Department of Physics \& Astronomy, McMaster University\\1280 Main St. W., Hamilton ON L8S 4M1, Canada.}


\date{\today}

\begin{abstract}
  The anisotropic triangular model, which is believed to describe the materials
  Cs$_2$CuCl$_4$ and Cs$_2$CuBr$_4$, among others, is dominated by
  incommensurate spiral physics and is thus extremely resistant to numerical
  analysis on small system sizes. In this paper we use twisted boundary
  conditions and exact diagonalization techniques to study the phase diagram of this
  model.  With these boundary conditions we are able to extract the inter- and
  intrachain ordering $q$-vectors for the $\frac{J'}{J} < 1$ region finding
  very close agreement with recent DMRG results on much larger systems.
  Our results suggest a phase
  transition between a long-range incommensurate spiral ordered phase, and a
  more subtle phase with short-range spiral correlations with the $q$-vector describing
  the incommensurate correlations varying smoothly through the transition.  
  In the latter phase correlations between next-nearest chains exhibits
  an {\it extremely} close 
  competition between predominantly antiferromagnetic and ferromagnetic correlations.
  Further analysis suggests that the antiferromagnetic
  next-nearest chain correlations may be slightly stronger than the
  ferromagnetic ones.  This difference is found to be slight but in line with
  previous renormalization group predictions of a collinear antiferromagnetic
  ordering in this region.  \end{abstract}


\maketitle

\begin{figure}[t]
\includegraphics[scale=0.5]{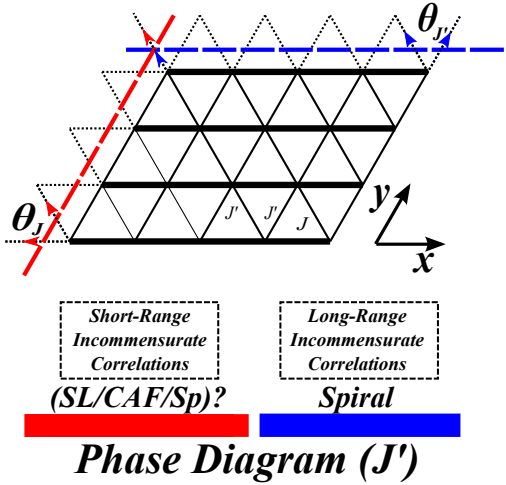}
\caption{\label{fig:Triangular_Lattice}  The anisotropic triangular lattice showing
a typical ED cluster.
The boundary twists
$\theta_J$ and $\theta_{J'}$, as discussed in the text, are as shown.  The colored arrows indicate the `positive' direction 
of the twist. The
upper leftmost bond receives a twist of $\theta_J + \theta_{J'}$. 
Below
the lattice diagram is a sample phase diagram showing an incommensurate spiral
ordering for $J'/J \sim 1$ with a transition to an unknown phase (speculated to
be either long-range collinear antiferromagnetically ordered (CAF),
short-ranged incommensurate spiral ordered or a two-dimensional spin liquid,
among other things).  
} 
\end{figure}

\section{Introduction}
\label{sec:Introduction}
In the study of two-dimensional quantum magnets, the anisotropic triangular
model has been a continuing object of attention.  This is partially due to its
applicability to real experimental materials such as the organic salts
$\kappa-$(BEDT-TTF)$_2$Cu$_2$(CN)$_3$,\cite{NMR_Organic_Salt,Sachdev_Organic_Salts,
  Organic_Salts_Hamiltonian_Work}
  $\kappa-$(BEDT-TTF)$_2$Cu$_2$[N(CN)$_2$],\cite{Organic_Salts_Hamiltonian_Work}
  and inorganic
  Cs$_2$CuCl$_4$,\cite{Coldea_Tsvelik_Neutron_Scatter_CsCuCl,Balents_Initial_CsCuCl,Coldea_Initial,Coldea_Hamiltonian_Work,Coldea_Zheng}
  and Cs$_2$CuBr$_4$,\cite{Coldea_Zheng,Tanaka_CsCuBr} and partially due to
  early theoretical and numerical speculation that it could exhibit a coveted
  2D spin liquid
  phase.\cite{Weng_Weng_Bursill_ED,Becca_Sorella_ED,Hauke_ED_MSWT,Chung_Marston_LSW_HATM}
  This was followed by suggestions that experimental results on
  Cs$_2$CuCl$_4$\cite{Coldea_Tsvelik_Neutron_Scatter_CsCuCl} could be explained
  by, less exotic, quasi-1D spin liquid
  behaviour.\cite{Kohno_Starykh_Balents_Nature,Balents_Nature}  This led to
  more recent theoretical work, utilizing renormalization group techniques,
  suggesting a subtle collinear antiferromagnetic (CAF) ordering in this same
  region; \cite{Balents_RG,Kallin_Ghamari_RG} this ordering being in
  competition with the more classical incommensurate spiral order, which also
  may exist.\cite{Pardini_and_Singh_Series_Expansion}  Most recently a DMRG
  study using periodic boundary conditions considered substantially larger
  systems than before and found a gapped state with strong antiferromagnetic
  correlations accented by weak, short-range, incommensurate spiral ones.\cite{Weichselbaum_and_White_DMRG}

Thus, the question in the $J' \ll  J$ region is whether the systems exhibits a
one- or two-dimensional spin liquid
phase,\cite{Becca_Sorella_ED,Weng_Weng_Bursill_ED,Hauke_ED_MSWT,Chung_Marston_LSW_HATM,Reuther_Thomale_Functional_RG}
or a collinear antiferromagnetic order driven by next-nearest chain
antiferromagnetic correlations and order by disorder,\cite{Balents_RG} or
something entirely different.  Suffice it to say that the true physics of this
system remains controversial.

Though Dzyaloshinskii-Moriya and interplane interaction are believed to play a
role in the physics of the previously mentioned real materials, the more
simplified system of a Heisenberg model on a triangular lattice with exchange
interactions $J$ along one direction and differing interactions ($J'$) along
the other two primitive vectors (see Fig.~\ref{fig:Triangular_Lattice}), is
believed to capture much of the relevant physics.  For $J'<J$ this can be
visualized as an array of weakly interacting chains.  In the limit of only two
chains this system reduces to the well studied $J_1-J_2$ chain, which is known
to be a gapless Luttinger liquid for $J \ll J'$ before undergoing a phase
transition at $J\simeq 0.24 J'$ to a gapped phase characterized by dimer-like and
incommensurate spiral
correlations.\cite{Thesberg_Sorensen_J1_J2_Chain,Chen_J1_J2,J1_J2_ED_1,Eggert,DMRG_2,DMRG_3,J1_J2_Field_Theory_1,Essler_TBC}
Though it is known that the behaviour of the true two-dimensional system
differs greatly.

In this paper we explore the $J' < J$ region of the anisotropic
triangular lattice Heisenberg model (ATLHM) through the use of twisted boundary conditions (TBC) and
exact diagonalization (ED).  This allows for a minimally biased
exploration of the incommensurate behaviour of the system. 
A typical cluster used in the calculations along with the imposed twists is shown in Fig.~\ref{fig:Triangular_Lattice}.
By minimizing the total energy of the ground-state with respect to the applied twist  we can determine the {\it optimal twist} $\theta^{gs}$ that most closely fit with the natural ordering
present in the system. It is then possible
to infer a preferred $q$-vector from the value of the $\theta^{gs}$. The inferred $q$-vector can tentatively be interpreted as the prefered $q$-vector for the system
in the thermodynamic limit. It is not limited to the usual discrete values $2\pi n/L$ but can take {\it any} value between 0 and $2\pi$. When such an analysis is performed
for the {\it ground-state} we can directly determine $q^{gs}$ for the ground-state, a substantial advantage of
the present approach.
We identify non-trivial values of $\theta^{gs}$  with the presence
of long-range spiral order.
Our results seem to indicate a phase transition between two gapless
phases: long-range spiral order with a non-trivial ground-state $q^{gs}\neq 0 $
and a more subtle phase with $q^{gs}=0$ and antiferromagnetic
intrachain ordering. At the critical point, the minimum in twist-space abruptly jumps between
two distinct minima resulting in a similar jump in the inferred $q^{gs}$. 
We very roughly estimate this transition to occur at a $J'_c \lesssim 0.5$ in the thermodynamic limit.  
However, we note that the severe limitation in system sizes when performing exact diagonalizations makes
it difficult to to draw a definitive conclusion concerning this transition in the thermodynamic limit.
The
interchain correlations of the latter phase are further explored with specific
attention paid to the competition between next-nearest chain antiferromagnetic
and ferromagnetic correlations as well as nearest chain incommensurate spiral
interactions. Our results, though not conclusive, seem to favor a CAF-like
ordering in this region. A schematic phase-diagram is shown in 
Fig.~\ref{fig:Triangular_Lattice}.

It is important to realize that the behavior of actual { correlation functions} are { not only }
determined by $q^{gs}$. In fact, 
following Ref.~\onlinecite{Essler_TBC}, we argue that the dominant part of the incommensurate transverse
correlations can be estimated by studying the {\it first excited state}. In general, 
$q$-vectors, describing the {\it transverse correlations}, are best determined by locating the twist
minimizing the energy of the first excited-state. If this minimum is located we can infer a $q^1$-vector
from which $q$, describing the incommensurate correlations, can be determined through the relation $q^1=q+q^{gs}$.
It is quite possible to have $q\neq 0$ and thus clear
incommensurate (short-range) correlations in the absence of long-range spiral order. Such short-range incommensurate would then
typically be modified by an exponentially decaying envelope.
Hence, by studying the minima of mainly the first excited-state,
we are able to extract the incommensurate $q$-vectors describing correlations along both the inter- and
intrachain directions. 
Our results for the intrachain $q$-vector  describing the incommensurate
correlations are in {\it very close}
agreement with recent DMRG results~\cite{Weichselbaum_and_White_DMRG} on substantially larger systems,
a strong validation of our approach. Further, the extracted $q$-vector for the correlations {\it varies smoothly} with $J'$ through the
tentative phase transition described above where $q^{gs}$ abruptly jumps showing that incommensurate correlations are
present on either side of the transition.

The organization of this paper is as follows: In section~\ref{sec:Introduction}
we introduce the model and its classical phase diagram, this is then followed
by an introduction to the twisted boundary conditions used here in
section~\ref{sec:Twisted_Boundary_Conditions} along with a detailed explanation of how $q^{gs}$ and $q$ are determined.  
We then show our results in
section~\ref{sec:Results_and_Discussion} along with analysis of the two phases.
We conclude in section~\ref{sec:Conclusion_and_Summary}.

\subsection{The Anisotropic Triangular Lattice Heisenberg Model (ATLHM)}

The system under consideration, the anisotropic triangular lattice Heisenberg
model (ATLHM), is described by the following Hamiltonian:
\begin{align}
\label{Hamiltonian}
H=J \sum_{\mathbf{x},\mathbf{y}} \hat{S}_{\mathbf{x},\mathbf{y}}\hat{S}_{\mathbf{x}-1,\mathbf{y}} + J' \sum_{\mathbf{x},\mathbf{y}} \hat{S}_{\mathbf{x},\mathbf{y}} \cdot \left( \hat{S}_{\mathbf{x},\mathbf{y+1}} + \hat{S}_{\mathbf{x}-1,\mathbf{y+1}} \right)
\end{align}
where for simplicity of exposition all lattice spacings $a$ are taken to be 1
and where $J>0$ corresponds to antiferromagnetic interactions.  A diagram can
be found in Fig.~\ref{fig:Triangular_Lattice}.  In this paper, we are solely
concerned with the $J' < J$ region, particularly the region where $J' \ll J$.
For reference, the anisotropy found in Cs$_2$CuCl$_4$ is estimated to be $J'/J
\sim 0.3$.\cite{Coldea_Tsvelik_Neutron_Scatter_CsCuCl} Throughout this paper we
use the convention that a system of size $N$ is composed of $W$ chains (i.e.
    width $W$) of length $L$ and is notated $N=W \times L$.

\subsection{The Classical System}

The classical limit case of the ATLHM (i.e. $S \rightarrow \infty$) can be
straightforwardly solved.\cite{Yoshimori_1959,Villain_1959} The lowest energy
configuration can be determined by positing a spiral solution of the form
$\mathbf{S} = S \mathbf{u} e^{-i \mathbf{q} \cdot \mathbf{r}}$.  This is
identical to a local rotation of the quantization direction at each site which
is done in spin-wave theory.  The resulting energy expression is then
\begin{align}
E_{cl}(\mathbf{q})= J \cos{(\mathbf{q}_{J})} + J' \cos{(\mathbf{q}_{J'})} + J' \cos{(\mathbf{q}_{J'} - \mathbf{q}_J)}
\label{eq:Ecl}
\end{align}
where the $ \hat{S}_i^z \hat{S}_j^z $ term is neglected since it carries no
$\mathbf{q}$ dependence.  For $J'<J$ we can find the minimum of Eq.~(\ref{eq:Ecl}) by first treating $\mathbf{q}_J$ as a fixed
parameter. In that case it immediately follows that the minimum with respect to $\mathbf{q}_{J'}$ is at
$2\mathbf{q}_{J'}=\mathbf{q}_J.  $
Thus, we get
\begin{align}
E_{cl}(\mathbf{q})= J \cos{(\mathbf{q}_J)} +2 J' \cos{\left( \frac{ \mathbf{q}_J}{2} \right)}.
\end{align}
The global minimum for $J'<J$ can now be found by minimizing this function with respect to $\mathbf{q}_J$. Solving with the
use of trigonometric identities yields the classical ground-state solutions:
\begin{align}
\mathbf{q}_J = 2 \arccos \left( -\frac{J'}{2J}\right), \; \mathbf{q}_{J'} =\arccos \left( -\frac{J'}{2J}\right).
\end{align}
However, for the region $J'/J =  ( 0 , 1 ] $ the $\mathbf{q}_J$ solution goes
from $\pi$ to $4 \pi /3$, we therefore choose a different solution
$\tilde{{\mathbf{q}}}_J = 2\pi - \mathbf{q}_J$, corresponding to a different
choice of branch, which ranges from the more physical $\pi$ to $2 \pi /3$.  The
$\mathbf{q}_{J'}$ solution needs no such adjustment.  Thus the final classical
solutions are
\begin{align}
q_J = 2\pi -2 \arccos \left( -\frac{J'}{2J}\right), \; q_{J'} =\arccos \left( -\frac{J'}{2J}\right).
\end{align}
where we no longer emphasize $q_J$ and $q_{J'}$ as vectors.  In the limit of
$J'/J \rightarrow 0$ we find $q_J=\pi$, $q_{J'}=\pi/2$, consistent with
antiferromagnetic chains with only perturbative coupling.

\section{Twisted Boundary Conditions}
\label{sec:Twisted_Boundary_Conditions}

The $J'\leq J$ region of the ATLHM is dominated by both incommensurate spiral
ordering and short-range incommensurate spiral correlations.  These
long-wavelength, incommensurate, correlations present formidable challenges to
numerical analysis, since attempts to capture physics with wavelengths of
$O(10,000) - O(\infty)$ using a system of length $\sim O(10)$ will undoubtedly
be dominated by extreme finite-size effects.  Even the most recent 2D DMRG
results, allowing for the largest systems, can only probe systems of $L \sim 100$
when at $J'=0.2$ the wavelength of the spiral correlations is expected to be on
the order of $10,000$.\cite{Weichselbaum_and_White_DMRG}  Thus, it is little
wonder that early numerical work produced such disputed
results.\cite{Weng_Weng_Bursill_ED,Becca_Sorella_ED}

Many of these finite-size effects can be successfully mitigated through a
careful consideration of the boundary conditions.  Previous numerical
studies\cite{Weng_Weng_Bursill_ED,Becca_Sorella_ED,Weichselbaum_and_White_DMRG}
on the ATLHM were produced using either open, periodic or mixed boundary
conditions.  Such boundary conditions will  strongly distort the physics of an
incommensurate system in favour of an ordering which is commensurate with the
system size, only admitting the ordering $q$-vectors
\begin{align}
q_n = \frac{2 \pi}{L} n \notag
\end{align}
where $L$ is the length of the system in a given direction.  It is this
tendency of them to ``lock'' a long wavelength structure into a much smaller
box that produces such spurious, unphysical, results such as sudden parity
transitions (a point to be discussed in greater detail below).\cite{Weng_Weng_Bursill_ED,Becca_Sorella_ED}
Thus to greatly reduce
this sort of error our calculations were performed using twisted boundary
conditions (TBC).

When using twisted boundary conditions, 
spin interactions which cross the periodic boundary of the otherwise translationally invariant system
become rotated in
the $x-y$ plane by an angle $\theta$.  This corresponds to the boundary
conditions
\begin{eqnarray}
S^-_{L+1}=e^{-i \theta} S^-_{1}, \;  S^+_{L+1}=e^{i \theta} S^+_{1}
\end{eqnarray}
or, equivalently,
\begin{eqnarray}
S^+_{L} S^-_1 \rightarrow S^+_{L} S^-_1 e^{-i\theta},\;S^-_{L} S^+_1 \rightarrow S^-_{L} S^+_1 e^{i\theta}.
\end{eqnarray}
where we simplify the discussion by only discussing one dimension of the
system.  Generalization to higher dimensions is straightforward although care has to be taken in order to
define positive and negative $\theta$ consistently when a twist is introduced along several bonds. (See Fig.~\ref{fig:Triangular_Lattice}. )
Physically, the twist corresponds to a spin current where a(n) $\uparrow$-spin
  ($\downarrow$-spin) acquires an extra phase when traversing the periodic
  boundary from the left (right).  Alternatively, the Heisenberg system can be
  mapped to one of $N_{\uparrow}$ fermions with the initial Jordan-Wigner
  transformation $S^+_i = c^{\dagger}_i e^{i \pi \sum_{i<j} c^{\dagger}_j
    c_j}$, followed by the gauge U(1) gauge transformation, $c^{\dagger}_i =
    f^{\dagger}_i e^{i\frac{\theta}{L}}$.  The interpretation is then of a
    periodic system, on a ring, of $N_{\uparrow}$ up spins threaded by a flux
    $\theta$.

\subsection{$J-J_2$ Spin Chain}
\begin{figure}[t]
\includegraphics[width=\linewidth]{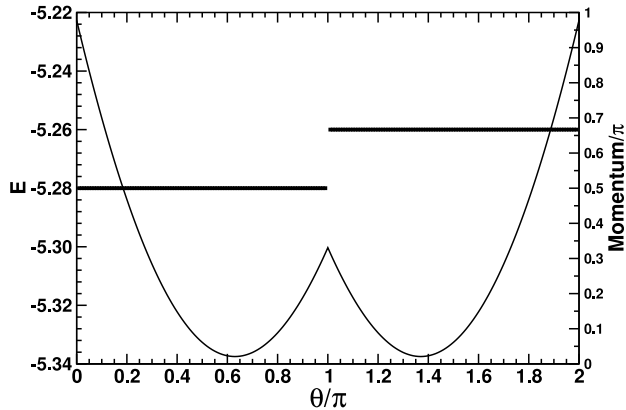}
\caption{\label{fig:EM12}  Energy and momentum of the lowest lying $S=1$ state for the
  $J-J_2$ chain at $J_2/J=1.$ The energy minima occur at $\theta=0.6299\pi,1.3701\pi$.
  At $\theta=\pi$ the lowest lying state changes from having $\tilde q = 6\pi/12$ to $\tilde q =8\pi/12$.
} 
\end{figure}

For an initially translationally invariant system of linear size $L$ a twist of $\theta$ imposed at the boundary
can then in general be distributed throughout the system by introducing a twist of $\theta/L$ at each bond by performing a non-unitary gauge-transformation.
We thereby obtain a model with periodic boundary conditions (PBC). Let us take
the well known $J-J_2$ spin chain model as an example:
\begin{align}
\label{J1J2}
H=J \sum_{i} \hat{S}_{i}\cdot\hat{S}_{i+1} + J_2 \sum_{i} \hat{S}_{i} \cdot \hat{S}_{i+2}.
\end{align}
This model is closely related to the ATLHM and was studied using twisted boundary conditions in
Ref.~\onlinecite{Essler_TBC} where a twist of $\theta$ was introduced at the boundary in the terms coupling sites $[L,1]$
as well as $[L-1,1]$ and $[L,2]$. We can in this case define a translationally invariant model with the {\it exact} same
energy spectrum if we instead introduce a twist of $\theta/L$ at each $[i,i+1]$ bond along with a twist of $2\theta/L$ at each
$[i,i+2]$ bond. This latter model is now manifestly translationally invariant with periodic boundary conditions and any many-body state can then be characterized
by a many-body momentum:
\begin{equation}
\tilde q =\frac{2\pi n}{L}\  \ n=0,1,\ldots,L-1
\end{equation}
To be explicit, if $T_a$ denotes the operator translating one lattice spacing $a$ in real space, then $T_a\Psi_{\mathrm{PBC}}=\exp(i\tilde q a)\Psi_{\mathrm{PBC}}$ with
$\Psi_{\mathrm{PBC}}$ the wave-function of the translationally invariant model with periodic boundary conditions.
We can then determine the energy as a function of $\theta$ as well as the many-body momentum of the corresponding state.  As an illustration,
results are shown in Fig.~\ref{fig:EM12} for the lowest lying $S=1$ excitation of the $J-J_2$ at $J=J_2$ for a chain with $L=12$,
        displaying the characteristic parabolic shape of the energy.
In this case the first energy minimum occurs at $\theta_{min}=0.6299\pi$ where $\tilde q=\pi/2$.
We then make the quasi-classical (phenomenological) assumption that the main effect of the twist is to modify the state's natural ordering
vector $q$ to fit with the many-body momentum $\tilde q$ in the following manner:
\begin{equation}
\tilde q=q\pm\frac{\theta}{L}=\frac{2\pi n}{L}.
\label{eq:q}
\end{equation}
In the present case we immediately find 
\begin{equation}
q=\pi/2+0.6299\pi/12.
\end{equation}
The second minimum at $\theta_{min}=2\pi-0.6299\pi$
and $\tilde q=2\pi/3$ yields the same 
\begin{equation}
q=2\pi/3-(2\pi-0.6299)/12=\pi/2+0.6299\pi/12.
\end{equation}
In the thermodynamic limit the natural ordering vector $q$ is then simply given by $\tilde q$ and any effects of the twist $\theta$ upon the dermination
of $q$ should be negligible as expressed by Eq.~(\ref{eq:q}).
This analysis differs in some details from Ref.~\onlinecite{Essler_TBC} but yields essentially identical results for the $J-J_2$ chain.

One may also consider the momentum of the model {\it without} translational invariance and twisted boundary conditions. In this case we find for the wave-function $\Psi_{\mathrm{TBC}}$
the relation $T_a\Psi_{\mathrm{TBC}}=\exp(i\alpha a)\Psi_{\mathrm{TBC}}$ with $\alpha=\tilde q+\theta N_\uparrow/L$ where $\tilde q$ is the many-body momentum
of the translationally invariant system. Here $N_\uparrow$ denotes the number of $\uparrow$ spins in the state under consideration.

\begin{figure}[t]
\includegraphics[scale=0.5]{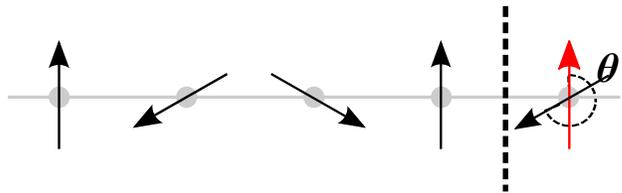}
\caption{\label{fig:Arrow_Diagram} 
This diagram shows how a $q=2 \pi/3$
  ordering can be made to ``fit'' into a system of length 4 by twisting by  $4
    \pi/3$ at the boundary.  These twisted boundary conditions then allow any
    incommensurate ordering to fit in any sized system. } 
\end{figure}

If one, at the classical level, argues that $\theta$ is the angle needed for $qL$ to equal an integer number of complete
turns one arrives at the same relation between $q$ and $\theta$:
\begin{align} 
q L \pm\theta =  2 \pi n,
\end{align}
In this equation, as well as in Eq.~(\ref{eq:q}), the $\pm$ signifies if $q$ turns in the same direction 
as $\theta$ as we move along the chain.
Hence the presence of the twist $\theta$
permit a continuum of ordering $q$-vectors 
to 'fit' into the system of linear size $L$, where:
\begin{align} 
q = \frac{1}{L} \left( 2 \pi n\pm \theta \right).
\label{eq:thetatoq}
\end{align}
A simple illustration of this is shown in Fig.~\ref{fig:Arrow_Diagram} for $q=2\pi/3$.
In this case the ordering can be made to ``fit" a system of length $L=4$ if a twist
$\theta=4\pi/3$ is introduced as indicated in Fig.~\ref{fig:Arrow_Diagram}. From $\theta$
we can then infer $q=(2\pi\times 2-4\pi/3)/4 = 2\pi/3.$ In this example, the wavelength
of the twist ($\lambda=3$) is shorter than the linear length of the system $L=4$ and we
have to use $n=2$ in Eq.~(\ref{eq:thetatoq}) in order to obtain the correct $q$. In analogy with
the example of the $J-J_2$ chain we would therefore expect the energy minimum for $\theta=4\pi/3$ to
occur for the state with many-body momentum $\tilde q=2\pi\times 2/4=\pi$. Correspondingly we would expect
another minimum at $\theta=2\pi/3$ for a state with many-body momentum $\tilde q=2\pi/4=\pi/2$. 

In practical studies it is not always feasible to use a translationally invariant system and explicitly determine
the many-body $\tilde q$ of the state corresponding to the minimizing twist and thereby $n$ in Eq.~(\ref{eq:q}) and (\ref{eq:thetatoq})
and for most of the results presented here we have not done so. However, it is almost always possible
to infer the correct $n$ to be used in Eq.~(\ref{eq:q}) and (\ref{eq:thetatoq}) by simple continuity from known results and
other expected behavior such as $qL\ll 1$.


\subsection{The ATLHM}
We now turn to a discussion of the approach we have taken to apply twisted boundary conditions to the
ATLHM.
With the analysis of the classical system in mind, we include \emph{two} twists in our
analysis of the ATLHM. The first, $\theta_J$, is associated with a
twisted boundary in the $J$ direction.  The second, $\theta_{J'}$, is then
associated with the boundary in the $J'$ direction (see Fig.~\ref{fig:Triangular_Lattice}).
With both twists implemented the Hamiltonian becomes:
\begin{align}
\label{Full_Hamiltonian}
&H_{\theta}=J\sum_{\mathbf{x}>1,\mathbf{y}} \hat{S}_{\mathbf{x},\mathbf{y}}\hat{S}_{\mathbf{x}-1,\mathbf{y}} + J' \sum_{\mathbf{x},\mathbf{y}<W} \hat{S}_{\mathbf{x},\mathbf{y}} \cdot \left( \hat{S}_{\mathbf{x},\mathbf{y+1}} + \hat{S}_{\mathbf{x}-1,\mathbf{y+1}} \right) \notag \\
&+  \sum_{\mathbf{y}<W}\hat{S}^+_{1,\mathbf{y}} \cdot \left( J \hat{S}^-_{L,\mathbf{y}} + J' \hat{S}^-_{L,\mathbf{y}}\right) e^{i \theta_J}  + J' \sum_{\mathbf{x}>1} \hat{S}^+_{\mathbf{x},W} \hat{S}^-_{\mathbf{x},1} e^{i \theta_{J'}} \notag \\
&+ J' \hat{S}^+_{1,W} \hat{S}^-_{L,1} e^{i(\theta_J + \theta_{J'})} + H.c. \notag \\
&+  \sum_{\mathbf{y}<W}\hat{S}^z_{1,\mathbf{y}} \cdot \left( J \hat{S}^z_{L,\mathbf{y}} + J' \hat{S}^z_{L,\mathbf{y}}\right) + J' \sum_{\mathbf{x}>1} \hat{S}^z_{\mathbf{x},W} \hat{S}^z_{\mathbf{x},1} e^{i \theta_{J'}}. \notag \\
\end{align}

Although this Hamiltonian looks quite cumbersome when written out explicitly,
conceptually it is very simple.  If a left moving $\downarrow$-spin traverses,
either horizontally or diagonally, the left periodic boundary it is rotated in
the $x-y$ plane by $\theta_J$.  If an upward moving $\downarrow$-spin
traverses, either vertically or diagonally, the upper periodic boundary it is
rotated in the $x-y$ plane by $\theta_{J'}$.  If a $\downarrow$-spin traverses
the upper left periodic boundary diagonally, thus crossing both twisted
boundaries, it is rotated in the $x-y$ plane by $(\theta_J + \theta_{J'})$.
Spins in the bulk as well as the $z$-component of all spins are unaffected by
the boundary.

These twists, which explicitly break the global SU(2) spin symmetry, are
identical to a twist of $\theta_J/L$ on \emph{each} horizontal and north-west to south-east bond
along with a twist of $\theta_{J'}/W$ on \emph{each} south-west to north-east and north-west to south-east bond.
The north-west to south-east bonds therefore recieve a twist of $\theta_{J}/L+\theta_{J'}/W$ for a system of dimensions
$W\times L$. (See Fig.~\ref{fig:Triangular_Lattice}.) If this is done one can work with an equivalent translationally invariant model.
However, a
twist-per-site approach was found to be less fruitful for such small systems
and the explicit SU(2) symmetry breaking will play an important role in forcing
a $S^z$ quantization direction which will be discussed below.

It is worth noting that the second twist, $\theta_{J'}$, is rarely (if ever)
implemented in studies with twisted boundary conditions.  Indeed, most existing
numerical studies of the ATLHM fail to consider the possibility of
incommensurate \emph{inter}chain correlations at all, often enforcing periodic
or open boundary conditions in the interchain direction even when other, more
elaborate, boundary conditions are used along chains.  The parameter
$\theta_{J'}$, then, serves as a tool to explore such new physics. 

Our complete Hamiltonian, boundary twists included, then has three free
parameters: The energy parameter $\frac{J'}{J}$, the intrachain boundary twist
$\theta_J$ and the interchain boundary twist $\theta_{J'}$.  The numerical task
then becomes to explore the two-dimensional landscape $(\theta_J,
    \theta_{J'})$, at a given $J'/J$, to find the twists which minimize the
ground-state energy.  From these twists the $q$-vectors $q_J$ and $q_{J'}$ can
then be extracted using the following generalization of 
Eq.~(\ref{eq:q}) and (\ref{eq:thetatoq}):
\begin{eqnarray}
\label{eq:2Dthetatoq}
q_J=\vec q_1\cdot \vec a_1 &=&\frac{2\pi  n_1}{L}\pm \frac{\theta_J}{L}\nonumber\\
q_{J'}=\vec q_2\cdot \vec a_2 &=&\frac{2\pi  n_2}{W}\pm \frac{\theta_{J'}}{W}.
\end{eqnarray}
These equations follow since the twists are applied in {\it direct} space and reflect the behavior of the system
upon $L$ and $W$ translations along the directions $\vec a_1$ and $\vec a_2$ in real direct space.
Our notation here for a system of $W$ chains of length $L$ is the following: As indicated in Fig.~\ref{fig:Triangular_Lattice} we
use basis vectors $\vec a_1=a(1,0)$ and $a_2=a(1/2,\sqrt{3}/2)$ for the direct lattice.
As usual reciprocal lattice vecors are then given by $\vec b_1=4\pi(\sqrt{3}/2,-1/2)/(a\sqrt{3})$ and
$b_2=4\pi(0,1)/(a\sqrt{3})$. If we now consider the translationally invariant model with twists
of $\theta_J/L$ and $\theta_{J'}/W$ along the bonds as described above, the many-body momentum of the
translationally invariant system with the imposed twist is:
\begin{equation}
\vec{\tilde q} = \frac{n_1}{L} \vec b_1+\frac{ n_2}{W}\vec b_2.
\end{equation}
Likewise, in our notation, we have: 
\begin{equation}
\vec q = \vec q_1+\vec q_2 = \frac{q_J}{2\pi}\vec b_1+\frac{q_{J'}}{2\pi}\vec b_2. 
\end{equation}
Hence, the application of
the twists allow us to determine the components of $\vec q$ along $\vec b_1$ and $\vec b_2$.

\begin{figure}[t]
\includegraphics[scale=0.5]{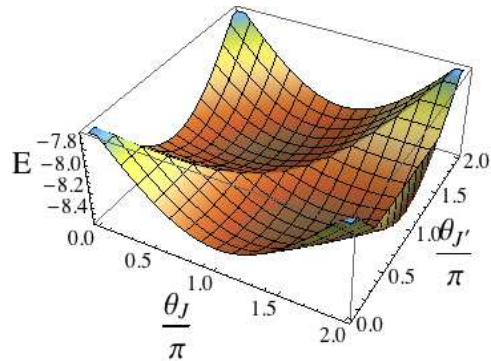}
\caption{\label{fig:Jp1sweep}  The energy, $E$, as a function of the 
two twists $\theta_J$ and $\theta_J'$. Results are shown for the lowest-lying $S=1$ state
of a $4\times4$ system with $J'/J=1$. The two identical minima occur for $(\theta_J,\theta_J')=(2\pi/3,4\pi/3)$ and $(4\pi/3,2\pi/3)$.
} 
\end{figure}

As an illustration we show in Fig.~\ref{fig:Jp1sweep} results for the $S=1$ ground-state energy of a $4\times 4$ system with $J'/J=1$.
Two identical minima are clearly present at $(\theta_J,\theta_J')=(2\pi/3,4\pi/3)$ and $(4\pi/3,2\pi/3)$. In this case we have done
simulations using a translationally invariant model as outlined aboved and explicitly determined the many-body momentum, $\vec{\tilde q}$,  of the state
corresponding to the minima. Here we find $(2\pi n_1/L,2\pi n_2/W)=(\pi/2,\pi)$ and $(\pi,\pi/2)$ respectively. Thus, following the analysis
at the end of the previous section we find $q_J=2\pi/3$. The mimima in the $S=0$ ground-state occur at the {\it exact} same $(\theta_J,\theta_{J'})$ but in this
case with $n_1=n_2=0$.

However, there is an additional complication in such an analysis brought on by
such small systems.  As has very clearly been shown in
Ref.~\onlinecite{Weichselbaum_and_White_DMRG}, much of the $J'/J < 1$ region is
dominated by \emph{antiferromagnetic} correlations superimposed on much subtler
incommensurate spiral correlations. Thus, if the system can be made to adopt a
specific quantization direction $z$, through, say, a perturbative magnetic
field on a single site as was done in
Ref.~\onlinecite{Weichselbaum_and_White_DMRG}, we expect $\langle GS \vert
\hat{S}_{\mathbf{x}}^z \hat{S}_{\mathbf{x+x'}}^z \vert GS \rangle$ correlations
to be completely dominated by antiferromagnetism with a small canted
incommensurate ordering showing in the transverse correlations:
\begin{align}
&\langle GS \vert  \hat{S}^x_{\mathbf{x}} \hat{S}^x_{\mathbf{x+x'}} + \hat{S}^y_{\mathbf{x}} \hat{S}^y_{\mathbf{x+x'}} \vert GS \rangle \notag \\
&\rightarrow \left\langle  \frac{1}{2} \left(  \hat{S}^+_{\mathbf{x}} \hat{S}^-_{\mathbf{x+x'}}+ \hat{S}^-_{\mathbf{x}} \hat{S}^+_{\mathbf{x+x'}}\right)\right\rangle&\propto \langle\hat{S}^+_{\mathbf{x}} \hat{S}^-_{\mathbf{x+x'}} \rangle.
\end{align}

In the absence of an explicit symmetry breaking term it is then extremely
difficult to separate the spiral correlations from the ``sea'' of
antiferromagnetic ones.  This difficulty is addressed by twisted boundary
conditions as can be seen through consideration of the following argument,
originally detailed and validated in Ref.~\onlinecite{Essler_TBC}.  First, with
the addition of a twist in the $x-y$ plane the global spin SU(2) symmetry is
broken and a unique $z$-quantization is picked out in a direction normal to the
system, since a generic twist would frustrate antiferromagnetic ordering
in-plane.  It is then convenient to rewrite the transverse correlations ,
  $\langle\hat{S}^+_{\mathbf{x}} \hat{S}^-_{\mathbf{x+x'}} \rangle$, in the
  more intuitive Fourier transformed form
\begin{align}
&=\left\langle\left( \frac{1}{\sqrt{L}} \sum_{q'} e^{i q x} \hat{S}^+_{q'}\right) \left( \frac{1}{\sqrt{L}} \sum_{q} e^{-i q (x+x')} \hat{S}^-_q\right) \right\rangle \notag \\
& = \frac{1}{L} \left\langle e^{-i q x'} \left( e^{i (q-q') x} \right)  \hat{S}^+_{q'} \left( \sum_m \vert m \rangle \langle m \vert \right) \hat{S}^-_{q}  \right\rangle \notag \\
&= \frac{1}{L} \sum_q \sum_{m} e^{-i q x'} \vert \langle m \vert S^-_q \vert GS \rangle \vert^2
\end{align}
where $S^-_q$ can now be physically interpreted as a spin-wave destruction
operator.  If the ground-state lies in the total $S^z=0$ sector, which it does
for an antiferromagnetic system of even system size, then $\langle GS \vert
S^-_q\vert GS \rangle=\langle GS \vert \left( \frac{1}{\sqrt{L}} \sum_q e^{-i q
    x} S_{\mathbf{x}}^-\right) \vert GS \rangle=0$ and the transverse
correlations can be rewritten as
\begin{align}
\langle\hat{S}^+_{\mathbf{x}} \hat{S}^-_{\mathbf{x+x'}} \rangle = \frac{1}{L} \sum_q \sum_{m \neq GS} e^{-i q r} \vert \langle m \vert S^-_q \vert GS \rangle \vert^2.
\end{align}

\begin{figure}[t]
\includegraphics[scale=0.5]{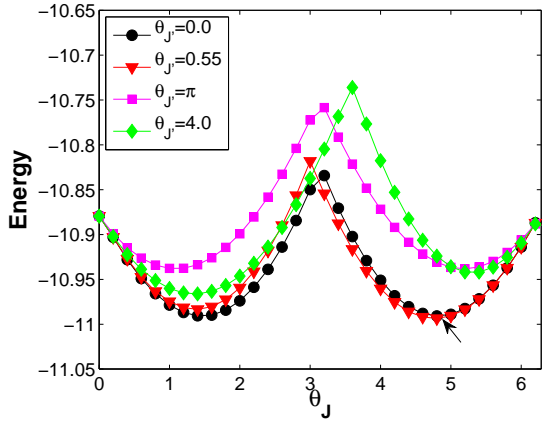}
\caption{\label{fig:Sample_Energy_vs_Theta}  
(Color online.) $Energy$ vs. $\theta_J$: The incommensurate and ground-state
  wavevectors $q_J^{in}$/$q_{J'}^{in}$ and $q_J^{gs}$/$q_{J'}^{gs}$ are
  obtained by minimizing the ground-state energy in the total-$S^z=1$ and
  total-$S^z=0$ subspaces respectively in terms of the boundary twists
  $\theta_J$ and $\theta_{J'}$.  This figure shows sample values of $\theta_J$
  vs. energy for different value of $\theta_{J'}$ for $N= 4 \times 6$ and
  $J'/J=0.6$ in the total-$S^z=1$ subspace.  The true minimizing $(\theta_{J}, \theta_{J'})$ were determined
  on a much finer grid to an accuracy of 0.001 in the twist, and for this case ($J'=0.6$) was found to be $(4.775, 0.550)$ (noted by an arrow) which corresponds to the $q$-vectors $(q_J, q_{J'})=(2.890,1.708)$.  This figure
  merely serves as an illustration. } 
  
\end{figure}

As usual, the $S^-_q$ or $S^-_x$ operators take the total $S^z=0$ ground-state
into the total $S^z=-1$ sector.  Additionally, if the ground-state has an
overall ordering vector $q^{gs}$ then the only terms to survive the sum over
$\langle m \vert S^-_q \vert GS \rangle$, and thus contribute to the transverse
correlations, are those for which $q^1=q^{gs}+q$.  If one then makes the
assumption that only the first excited state in the total $S^z=1$ sector
dominates then one now has a method to extract the incommensurate $q$-vector,
$q$, as well as the ground-state momentum $q^{gs}$.  First one
finds the $(\theta_J, \theta_{J'})$ which minimizes the ground-state
energy of the total $S^z=0$ sector, yielding
\begin{equation}
(q^{gs}_J,q^{gs}_{J'}) \ \ (S^z=0).
\end{equation}
Notice that our two twists, $\theta_J$ and
$\theta_{J'}$ yield two $q$-vectors which we denote $q_J$ and $q_{J'}$. 
After
finding the minimum in the total $S^z=0$ twist-space the procedure is then
repeated in the total $S^z=1$ twist-space
yielding 
\begin{equation}
q^1_J=q^{gs}_J+q_J\  \mathrm{and}\ \ q^1_{J'}=q^{gs}_{J'}+q_{J'}\ \ (S^z=1).
\end{equation}
A demonstration of
this can be found in Ref.~\onlinecite{Essler_TBC}. 
As an illustration of the procedure we show
results for $E(\theta_J,\theta_{J'})$ for the
first excited state of a $4\times 6$ system with $J'/J=0.6$ in Fig.~\ref{fig:Sample_Energy_vs_Theta}. Note that, for our subsequent results the minima
are determined on a much finer grid. We also note that in both Fig.~\ref{fig:Jp1sweep} and \ref{fig:Sample_Energy_vs_Theta} do distinct minima occur for
values of $\theta>\pi$. This is due to the non-zero $\theta_{J'}$ which lifts the symmetry with respect to $\theta=\pi$ visible in Fig.~\ref{fig:EM12}.

This method of minimizing the ground-state energy in both the total $S^z=0$ and
$S^z=1$ sectors, also allows one to compute the \emph{spin gap},
$\Delta$, between these states.  Thus, with knowledge of the spin gap, the
ground-state long-range ordering $q$-vectors as well as the incommensurate
short range $q$-vectors, one can imagine
two situations of interest that could arise in the ATLHM. 

\subsection{Case 1: $q^{gs} \neq 0$ or $\pi$, $q = 0$ (incommensurate spiral order)}
\label{subsec:Case_1}

In the case where the true ground-state (i.e. that in the total $S^z=0$ sector)
is minimized by incommensurate $q$-vectors $q^{gs}_J$ and $q^{gs}_{J'}$ we then
have incommensurate long-range order related to a classical incommensurate
spiral. 

In such a region we also expect the spin gap ($\Delta$) to vanish owing to the
gapless magnon excitations about the spiral order which accompany U(1) symmetry
breaking.  Note that the symmetry broken \emph{is} U(1), since the initial
SU(2) symmetry has already been reduced to U(1) when the twist terms were
added.  This would coincide, in the limit of infinite system size, with
long-range correlations of the form
\begin{align}
\left\langle \hat{S}_{\mathbf{x},\mathbf{y}}\hat{S}_{\mathbf{x} + \mathbf{x'},\mathbf{y}}\right\rangle \underset{x' \rightarrow \infty}{\approx} e^{i q_{J}^{gs} x' } \left\langle \hat{S}_{\mathbf{x}}\right\rangle^2,
\end{align}
with a similar form in the $J'$ direction corresponding to $q_{J'}^{gs}$.
However, such long-range behaviour of the correlation functions is far beyond
the accessible range of any numerical approach.  Thus, it will suffice to take a
non-zero $q^{gs}$ accompanied by a vanishing spin gap $\Delta$ to demonstrate
long-range spiral order.

\subsection{Case 2: $q^{gs}=0$ or $\pi$, $q \neq 0$ (non-spiral order)}
\label{subsec:Case_2}

The case where $q^{gs}=0$ or $\pi$ is more 
complicated.  Since $q$ is non-zero the system is displaying
incommensurate spiral correlations, however, these correlations are of
insufficient strength to stabilize true long-range spiral ordering.  Yet, as we
find, if the spin gap $\Delta$ is found to be zero, then we expect \emph{some}
ordering to exist, unless the system is found to be a gapless spin liquid.  

\begin{figure}[t]
\includegraphics[scale=0.5]{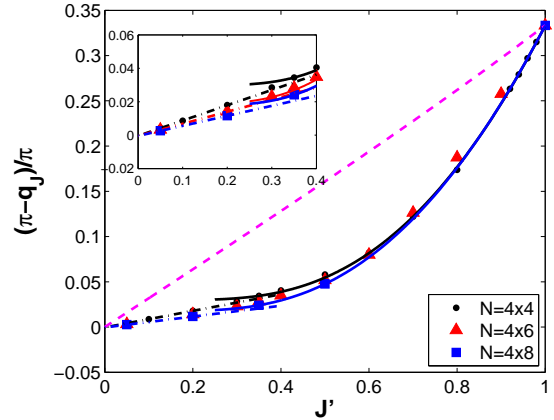}
\caption{\label{fig:q_J_incommensurate_vs_Jp}  (Color online.) $q_J$ vs. $J'$:
The intrachain ordering $q$-vector $q_J$ as a function of the interchain
  interaction $J'$ for systems of width 4 along with the classical value
  (dashed magenta line).  
  Results are obtained from the 
  $\theta_{J}$ which minimizes the total-$S^z=0,1$ sectors.
  For $J'>J'_c$ $q_J^{gs}$ was found to be non-zero, while $q_J^{gs}=0$ or $\pi$ for $J'<J'_c$
  The critical
  value, $J'_c$ was determined to be $J'_c=$0.9175, 0.7835, 0.7135 for $N= 4 \times 4$, $4
 \times 6$ and $4 \times 8$ respectively.  
  Exponential fits (black for $N=4
      \times 4$, light grey for $N= 4 \times 6$ and dark grey for $N = 4 \times
      8$) are of the form $a (J')^2 \exp(-b/J')$, consistent with
  Ref.~\onlinecite{Weichselbaum_and_White_DMRG}, and are found to be extremely
  good for most of the $J'$ region.  However at $J' \sim 0.3$ the data markedly
  deviates from this fit and develops a linear character.  The physicality of
  this linear behaviour for $J' < 0.3$ is further explored in the text. 
}
  \end{figure}

\begin{figure}[t]
\includegraphics[scale=0.5]{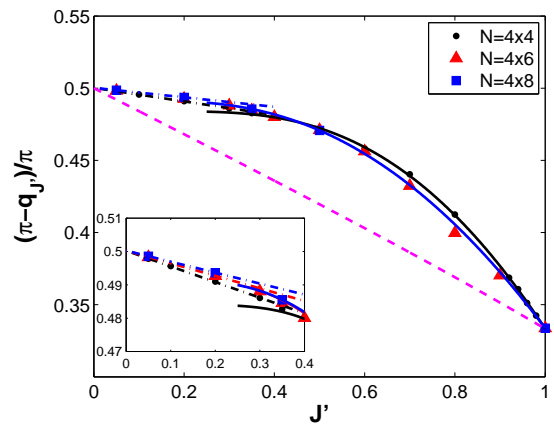}
\caption{\label{fig:q_Jp_incommensurate_vs_Jp}  (Color online.) $q_{J'}$ vs. $J'$:
The interchain ordering $q$-vector $q_{J'}$ as a function of the interchain
  interaction $J'$ for systems of width 4.   
  Behaviour and fits are identical to those in
  Fig.~\ref{fig:q_J_incommensurate_vs_Jp}.  As $J' \rightarrow 0$,  $q_{J'}$  }
\end{figure}

Should the value of $q^{gs}$ be consistent with a $\pi$ ordering vector for all
system sizes then, taken along with the gaplessness of the system, one could
conclude that the system has ordered antiferromagnetically.  However, in our
case a more thorough
analysis of the correlations of the system becomes necessary.  This is expanded
upon in Sec.~\ref{subsec:Non_Spiral_Ordered_Phase}

\section{Results and Discussion}
\label{sec:Results_and_Discussion}

The intrachain ($\theta_J$) and interchain ($\theta_{J'}$) boundary twists were
varied to minimize the ``ground-state'' energy in the total-$S^z=0,1$ sectors for
systems of increasing length and fixed width (4 chains).  A fixed width was
chosen both since intrachain correlations are the dominant correlations for $J'
\ll J$ and to more easily compare with existing DMRG work on larger
systems.\cite{Weichselbaum_and_White_DMRG}  From these minimizing twist values
the $q$-vectors $q_J$ and $q_{J'}$ were extracted as a function of $J'/J$ which we will simply call $J'$ (i.e. $J=1$).
The resulting data, as well as
the classical values, can be found in Fig. \ref{fig:q_J_incommensurate_vs_Jp}
for $q_J$ and Fig. \ref{fig:q_Jp_incommensurate_vs_Jp} for $q_{J'}$.  With the
exception of the $J' \lesssim 0.3$ region which is discussed later, both $q_J$
and $q_{J'}$ are found to be fitted best by functions of the form $a (J')^2
\exp(-b/J')+c$ rather than power-law fits.  This data is in close agreement
with the DMRG results found in Ref. \onlinecite{Weichselbaum_and_White_DMRG}
where the incommensurate $q$-vector $q_{J}$ ($q_{J'}$ was not considered) was
extracted by fitting $\langle S^z_{\mathbf{x}}\rangle$, as induced by a boundary field, to an exponentially decaying correlation function of the
form 
$\left\langle S_{\mathbf{0}}^z\right\rangle \exp(-x/\xi)
  \cos(q x)$.  The close agreement of our results with the DMRG results on
  substantially larger systems is surprising and indicative of the power of
  twisted boundary conditions to circumvent finite-size effects in
  incommensurate systems.

\begin{figure}[t]
\includegraphics[scale=0.45]{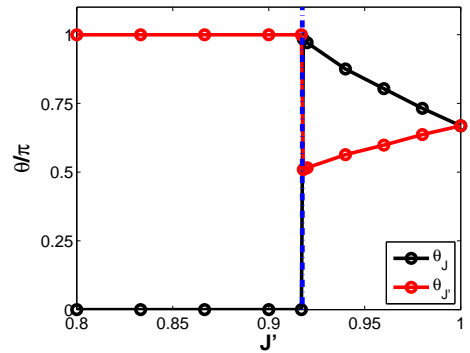}
\caption{\label{fig:Sample_L16_Sz_0_Jump}  (Color online.) 
 $\theta_J$, $\theta_{J'}$ vs. $J'$ in
the total $S^z=0$ subspace ($N=4 \times 4$): As can be clearly seen the
total $S^z=0$ minimizing boundary twists (shown as solid lines with circular
    markers) undergoes an abrupt jump (occurring here
        at $J'/J \sim 0.91$) before ``locking'' to some fixed value for all
      $J'<J'_c$. 
}
\end{figure}

The lack of distinct features in Fig. \ref{fig:q_J_incommensurate_vs_Jp} and
Fig. \ref{fig:q_Jp_incommensurate_vs_Jp} suggests that the system has identical
behaviour for all $J'$. This is not the case, as can be seen by examining the
true ground-state $q$-vectors in the total-$S^z=0$ sector.  As $J'$ decreases
the ground-state twists, $\theta^{gs}_J$ and $\theta^{gs}_{J'}$, are found to jump
discontinuously at some critical value of $J'$, $J'_c$ (the nature of this jump
    will be discussed momentarily).  For $J' \> J'_c$ the total-$S^z=0$ and
total-$S^z=1$ twists coincide.
Below $J'_c$ the
total-$S^z=0$ data is found to be either $0$ or $\pi$ for all $J'$ in the
region.  
A sample illustration of this jump can be found in
Fig.~\ref{fig:Sample_L16_Sz_0_Jump} where it is shown for a $4 \times 4$
system.  This critical value decreases with system size and is found to be at
$J'_c=$0.9175, 0.7835, 0.7135 for $N= 4 \times 4$, $4 \times 6$ and $4 \times
8$ respectively.  A finite-size extrapolation of these values to the
thermodynamic limit can be found in Fig. \ref{fig:Jp_Critical_vs_N}. 
At $J'_c=0.9175$ the $\theta^{gs}_J$ minimizing the energy jumps abruptly to $0$ due
to the appearance of a new distinct minimum in twist space.
We can extrapolate $J_c'$ to the thermodynamic limit with a linear $\frac{a}{N}+b$ fit of $N^{-1}$ estimating
$J'_c\to 0.475$  as $N\to\infty$.
As the system width is
increased $J'_c$ is found to increase as well, taking values of 0.912 and 0.917
for systems of size $N= 6 \times 4$ and $N=8 \times 4$.  The infinite system
size extrapolation for fixed length, which is obviously non-linear and is thus fitted with a quadratic $\frac{a}{N^2} +
\frac{b}{N}+c$ fit, can be
found in the inset of \ref{fig:Jp_Critical_vs_N} and is found to be 0.948. 
Obviously, for these very limited system sizes, a reliable estimate of the critical coupling
in the thermodynamic limit is not within reach. However, it seems plausible that the fixed
width estimate of $J'_c=0.475$ is the more realistic of our estimates.
A comparison of both
fixed width and fixed length thermodynamic limit extrapolations suggests that
the spiral-ordered region extends well into the $J'<J$ region, even in much
larger systems.

It is important to note that this discontinuous jump in the ground-state
is \emph{not} due to the level crossing observed in previous numerical
work.\cite{Weng_Weng_Bursill_ED,Becca_Sorella_ED}  This transition, which was
found to be a parity transition, occur for a $4 \times
4$ system at a value of $J' \sim 0.84$, for $4 \times 6$ and at $J' \sim 0.75$ for
$4 \times 8$ and thus occurs at a higher value of $J'$ for all system sizes.
Thus, this level crossing is completely avoided once one allows the boundaries
to twist freely.

The nature of the transition is essentially due to a first-order phase
transition in ``twist-space'' as described by Landau theory. At $J' > J'_c$ the
ground-state minima is found to lie at some incommensurate twist value, at $J'
\sim J'_c$ a second commensurate minimum forms elsewhere, at say
$(\theta_J,\theta_{J'})=(0,\pi)$, this second minima then lowers in energy as
the incommensurate minima, which is still the global minima, rises.  At
$J'=J'_c$ the commensurate minimum overtakes the incommensurate one to
become the new global minimum and the ground-state then jumps
discontinuously.

\begin{figure}[t]
\includegraphics[width=0.55\textwidth]{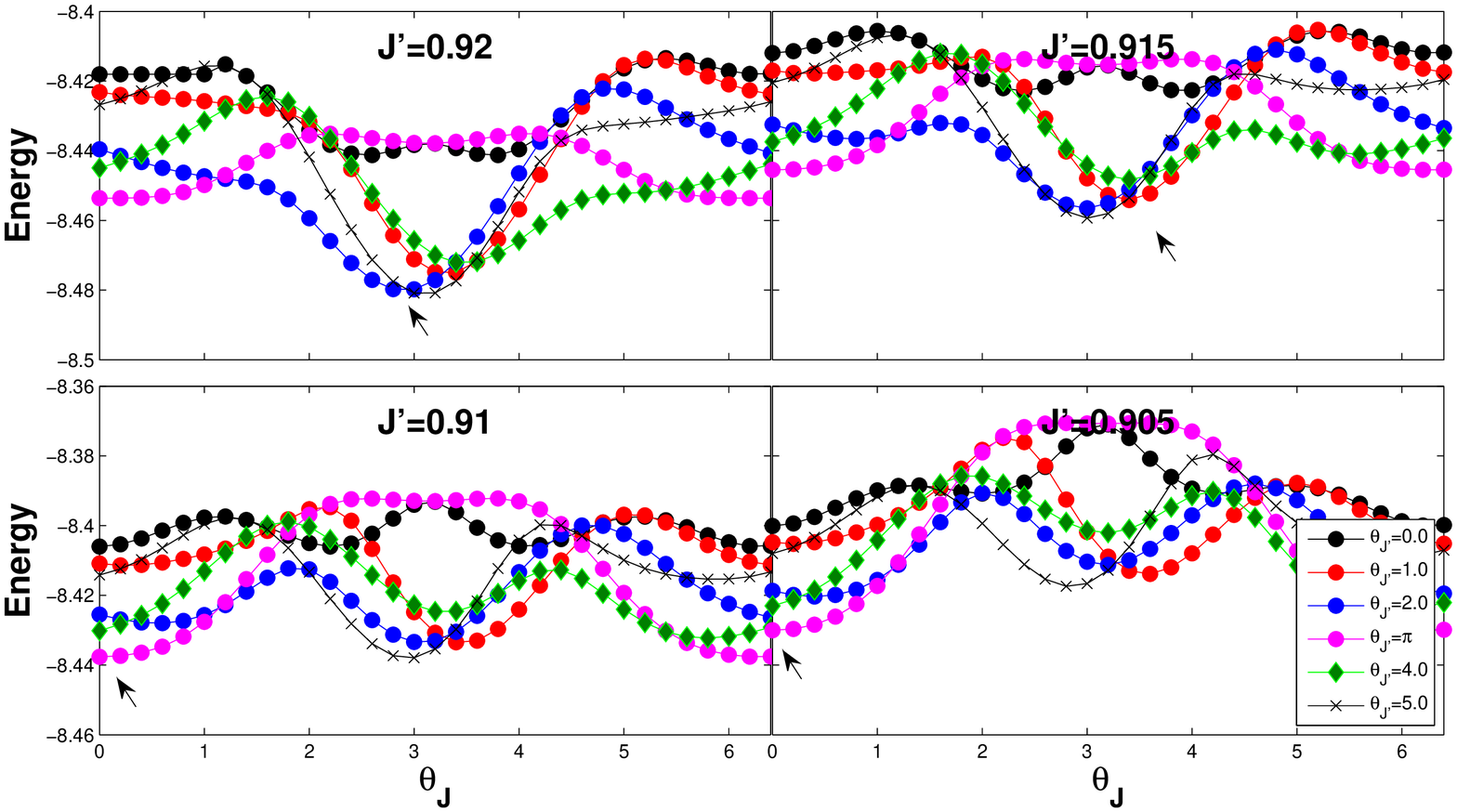}
\caption{\label{fig:Theta_vs_Energy_Transitions_Graphs}  (Color online.) 
Energy vs. $\theta_J$ for varying values of $\theta_{J'}$ shown for different
  values of $J'$ near $J'_c=0.915$ in the total $S^z=0$ subspace ($N=4 \times
      4$). At $J' > J'_c$ the ground-state minima is found to lie at an
  incommensurate twist value, at $J' \sim J'_c$ a second commensurate minimum forms at
  $(\theta_J,\theta_{J'})=(0,\pi)$, this second minimum then moves lower in energy 
  and becomes the global minimum at $J'=J'_c$.  The global minima is indicated in the graphs with an arrow.}
\end{figure}

We now look at each region separately:

\begin{figure}[t]
\includegraphics[scale=0.5]{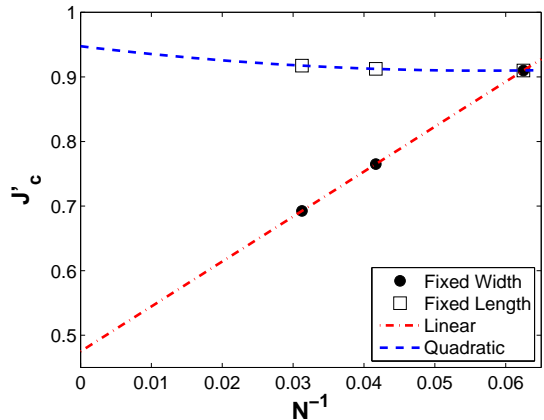}
\caption{\label{fig:Jp_Critical_vs_N}  (Color online.) Critical $J'$ vs $N^{-1}$:
  A thermodynamic limit extrapolation of the spiral-ordered to non-spiral
    ordered transition value ($J'_c$) for systems of width 4 and length 4.  For
    a width of 4 a linear ($J'_c = 0.475$ as $N\rightarrow \infty$) fit was considered.
    For a length of 4 the extrapolation was clearly not linear and so a
    quadratic fit ($J'_c=0.948$ as $N\rightarrow \infty$) was used.   For
    consistency the quadratic extrapolation values for both analyses are
    considered the best fit. The critical temperature was indicated by a
    discontinuous jump in $\theta_J$ and $\theta_{J'}$ from their $J' \ll 1$
    values.} \end{figure}

\subsection{The Incommensurate Spiral Ordered Phase, $1 \geq J'/J > J'_c/J$}
For the isotropic case, where $J' =1$, the ordering $q$-vectors
$(q_J,q_{J'})$ were found to be $\left( 2 \pi/3 , 2 \pi/3 \right)$ in agreement
with previous work.  As $J'$ decreases the $q$-vectors then vary continuously
through incommensurate values.  In this region the energy minima of the
total-$S^z=0$ and total-$S^z=1$ regions coincide in twist space.  This, taken with the lack of
an energy gap (this is shown in section \ref{sec:The_Energy_Gap}), indicates
that this region is in a long-range spiral order phase.  The transition out of
this phase seems to occur at an intrachain twist of $\sim \pi$ for all system
sizes as illustrated in Fig.~\ref{fig:Sample_L16_Sz_0_Jump} for a $4\times 4$ system.
The fact that the transition should be related to some critical value
of the boundary twist \emph{and not} some critical $q_J$ is interesting and may
represent some subtle numerical cause.  However, we would comment that spin
wave theory is known to encounter a similar region, notable for its
non-convergence, for $J'$ smaller than some critical
value\cite{Hauke_ED_MSWT,Chung_Marston_LSW_HATM}. On the other hand, we also cannot exclude
the possibility that for much larger systems this transition would be absent.

\subsection{The Non-Spiral Ordered Phase, $J' < J'_c$}
\label{subsec:Non_Spiral_Ordered_Phase}

For $J'$ values greater than $J'_c$ the ground-state is found to have
incommensurate long-range spiral order as was discussed previously.  However,
for $J'<J'_c$ the twists which minimize the total $S^z=0$ sector jumps to
$(\theta_{J},\theta_{J'})=(0,\pi)$ for $N=4\times4$, $4 \times 6$ and $4 \times
8$ (i.e. systems of width 4), and to $(0,0)$ for systems of size $N=6 \times 4$
and $N=8 \times 4$.  These values of $\theta_J$ are found to be entirely
consistent with antiferromagnetic intrachain ordering of the ground-state.
However, for increasing system width, the values of $\theta_{J'}$, being $\pi$
for width 4 but $0$ for widths of both 6 \emph{and} 8, are inconsistent with
any $q$-vector suggesting a more careful consideration of interchain physics
must be taken.  This discussion is postponed until section
\ref{subsec:Interchain_Correlations}.

The fact that no evidence of this transition can be found in the total $S^z=1$
data suggests that incommensurate correlations are always present and vary
smoothly for all $1>J'/J>0$, but that the power of those correlations to
stabilize long-range spiral order becomes insufficient at $J'_c$.  Below $J'_c$
the dominant correlations are then antiferromagnetic along chains with much
smaller incommensurate behaviour resting atop.  These antiferromagnetic
correlations nested in an incommensurate envelope were demonstrated very
clearly for a gapped system in Ref.~\onlinecite{Weichselbaum_and_White_DMRG}.
Though it is our belief that this behaviour is only found below $J'_c$ and the
fact that such behaviour was obtained for $1 \geq J'/J > J'_c$ in that paper
might be an artefact of the use periodic boundary conditions in the interchain
direction there.  This point is further discussed in the next subsection.

\begin{figure}[t]
\includegraphics[scale=0.5]{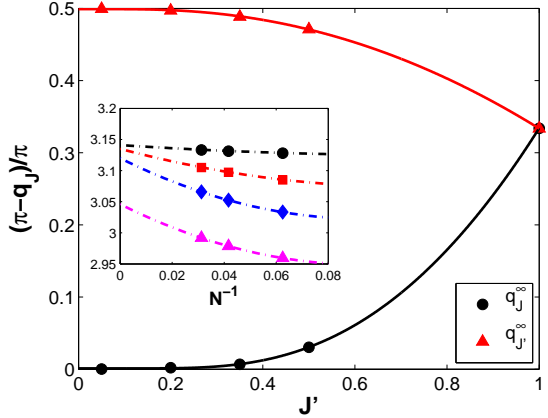}
\caption{\label{fig:infinite_q_incommensurate_vs_Jp}  (Color online.) $q^{\infty}_J$ and $q^{\infty}_{J'}$ vs. $J'$:
The thermodynamic limit extrapolated values of $q_{J}$ and $q_{J'}$ vs. $J'$.
  Extrapolations were done to quadratic functions of the form $\frac{a}{N^2} +
  \frac{b}{N} + c$.  Sample extrapolations can be seen in the inset for
  $J'/J=0.5$ (triangles), $0.35$ (diamonds), $0.2$ (squares) and $0.05$
  (circles).  Values greater than the estimated infinite system size transition
  point and less than $J'/J=1$ are not shown (see text).  In similarity to the
  finite-size Fig.~\ref{fig:q_J_incommensurate_vs_Jp} and
  Fig.~\ref{fig:q_Jp_incommensurate_vs_Jp}, $q_J\rightarrow \pi$ and $q_{J'}
\rightarrow \pi/2$ as $J' \rightarrow 0$.  However, contrary to that figure,
  the degradation of an exponential fit to a linear one in the $J' \sim 0$
    region is less pronounced, if present at all (see text).} 
\end{figure}

Numerical access to three system sizes of width 4 makes it possible
to extrapolate $q_J$ and $q_{J'}$ to the $4 \times \infty$ limit.  Extrapolated
values were found to lie, with great precision, on a scaling function of the
form $\frac{a}{N^2} + \frac{b}{N} + c$ and can be found in the inset of
Fig.~\ref{fig:infinite_q_incommensurate_vs_Jp}.  The thermodynamic limit
results for both $q^{\infty}_J$ and $q^{\infty}_{J'}$ are plotted in the main
figure.  Values above $J'_c$ were not considered, with the exception of the
commensurate $J'/J=1$ case. As before, these $q^{\infty}_J$ and
$q^{\infty}_{J'}$ data can be well fitted by a function of the form $a (J')^2
exp(-b/J')+c$.  However, unlike the finite-size case, this function is found to
be valid for all $J'$ considered.  This suggests that the linear behaviour in
the neighbourhood of $J' \sim 0$ (See Fig.~\ref{fig:q_J_incommensurate_vs_Jp}) may not be physical.  Furthermore, as will be
discussed in section \ref{subsec:Interchain_Correlations}, it is found for $J'
\lesssim 0.3$ that the system's energy dependence on $\theta_{J'}$ becomes zero
to numerical precision.  Thus it is possible that interchain correlations,
which are physically non-zero, but of a magnitude smaller than the smallest
number that could be represented by a computer are present in this region.
Regardless, the physicality of the linear behaviour is not certain.

\subsubsection{The Energy Gap: $\Delta$}
\label{sec:The_Energy_Gap}

The numerical determination of a spin gap is in general a difficult task.
Often computational reality doesn't permit enough system sizes to be calculated
in order for a reliable thermodynamic limit to be established.  Furthermore,
when the thermodynamic limit can be taken, considerations like the method
and boundary conditions used can have a profound effect on the extrapolated
results.

\begin{figure}[t]
\includegraphics[scale=0.5]{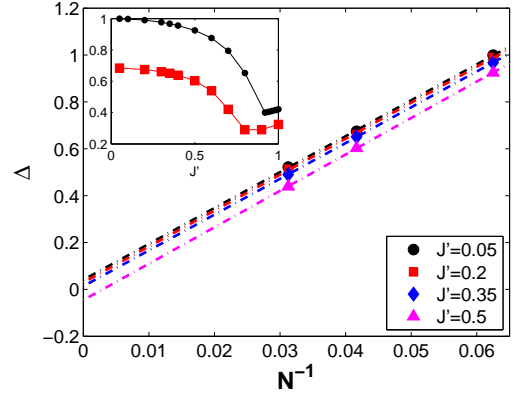}
\caption{\label{fig:Energy_Gap_vs_oN}  (Color online.) Spin Gap vs. $N^{-1}$:
The energy difference between the ground-states of the total $S^z=0$ and total
  $S^z=1$ sectors extrapolated to the thermodynamic limit.  Dotted lines
  represent fits of the form $\frac{a}{N}+\Delta_{\infty}$ with
  $\Delta_{\infty}$s found in the non-spiral region to be on the order of
  $10^{-2}$.  The error, estimated by contrasting fits quadratic vs. linear in
  $N^{-1}$, could be as large as $10^{-1}$; or approximately one percent.
  Inset: $\Delta$ vs. $J'$:  We show the energy gap $\Delta$ versus $J'$ for
  $N=4\times 4$ (circles) and $N=4 \times 6$ (squares).  Without a
  thermodynamic limit analysis it is easy to see how previous work found the
  $J'<J'_c$ region to be gapped.} \end{figure}

With this in mind the bulk of existing numerical work on the ATLHM has
suggested the existence of a spin gap either for all of $J'/J < 1$ or
for $J'$ less than some critical value in the range of $J'/J \sim 0.6 -
0.8$.\cite{Becca_Sorella_ED,Weng_Weng_Bursill_ED,Weichselbaum_and_White_DMRG}
Indeed, an initial analysis of our own data, as can be seen in the
inset of Fig.~\ref{fig:Energy_Gap_vs_oN}, is consistent with this
picture.  However, a more careful consideration of these results shows that this could be misleading.
	
As previously, the accessibility of three width 4 system sizes permits
a finite-size extrapolation of the energy gap data.  This extrapolation
was done for values of $J' \leq 0.5$ and can be found in
Fig.~\ref{fig:Energy_Gap_vs_oN}.	 Values of $1 > J' \geq 0.5$ were not
considered due to the possibility of different system sizes being on opposite
sides of the $J'_c$ transition.  For $J' \leq 0.5$ the $\Delta$ values were
found to fit very well to a scaling function of form
$\frac{a}{N}+\Delta_{\infty}$ and $\Delta_{\infty}$ was found to be on the
order of $10^{-2}$.  An estimate of the error in this extrapolation can be
generated by contrasting the linear fit y-intercept with that of a quadratic
fit which produces a $\Delta_{\infty}$ on the order of $10^{-1}$.  Such small
values are extremely suggestive of a gapless system.  Taken alone, this is
consistent with both spiral and collinear antiferromagnetic orderings as well
as potentially a gapless spin liquid phase.
	
Data was also collected for $J'/J=1$ where the situation appeared to be
different, with linear scaling fits suggesting a small non-zero spin gap.
However, it is well
known\cite{Trumper_Sorella_Capriotti_VMC_Spin_Gap_Extrapolation,Trumper_Sorella_Capriotti_VMC_Spin_Gap_Extrapolation_2}
that the spin gap converges very slowly with system size in the spiral-ordered
region and the system is known to be gapless in this phase.  This, combined
with the apparent gaplessness of the $J'<J'_c$ region, suggests that the ATLHM
is gapless for all $J'/J \leq 1$

One of the central results of the numerics of
Ref.~\onlinecite{Weichselbaum_and_White_DMRG} was the unusual behaviour of the
energy gap  $\Delta$ for different system widths.  This paper studied the
incommensurate behaviour of long systems of a small number of chains, i.e. $4
\times 64$, $6 \times 64$, $8 \times 32$, etc.  One of the central results of
the paper was that for systems (periodic in $\mathbf{y}$) of width 2, 4 and 8
the system exhibited a spin gap for $J'<1$ which shrank with decreasing $J'$
down to $J'\sim 0.5$, the lowest $J'$ studied in the work.  This spin gap was
accompanied by an exponential decay of intrachain correlations.  Conversely,
systems of width 6 displayed a small or, likely, no such spin gap and
presumably an algebraic decay of correlations.  The reason for this discrepancy
could not be identified.  A possible explanation for the discrepancy between these
results and the ones presented here could be the presence of a non-zero $\theta_{J'}$ in
our calculations allowing the system to relax more completely as we now comment on in more detail.

For $J'/J$ in the neighbourhood of $1$ the classical and quantum ordering
vector in the $\mathbf{y}$ direction is $q_{J'}=2 \pi /3$, where for $J'/J \ll
1$ $q_{J'} \rightarrow \pi/2$ as $J' \rightarrow 0$.  Thus, a cylinder, as used in Ref. \onlinecite{Weichselbaum_and_White_DMRG}, with a
width of 6 chains and periodic (no twist) boundary conditions around the cylinder would be commensurate with
the $2 \pi/3$ order but not the $\pi/2$ order, and thus we expect the correct
spin gap for $J'/J \sim 1$ and an artificial, finite size induced gap as $J'$
decreases (though this is not observed in Ref.  \onlinecite{Weichselbaum_and_White_DMRG}
since the spin gap is only calculated as low as $J' \sim 0.9$ for the $6 \times 64$ system).
Conversely, system sizes of 4 and 8 are incommensurate with $2 \pi/3$ order,
  and therefore are found to have an unphysical spin gap when $J'\sim J$, but
  \emph{are} commensurate with a $q_{J'}=\pi/2$ ordering and thus we expect the
  correct spin gap to emerge as $J' \rightarrow 0$.  It is then the case that
  the 4 and 8 width system would be expected to give the most accurate
  indication of the spin gap for small $J'$ and the width 6 system for $J'/J
  \sim 1$.  The key point is that gapless spiral correlations in the $J'$
  direction might appear gapped if analyzed with periodic boundary conditions around the cylinder
  with widths incommensurate with the spiral in that direction. With this in mind, an alternative interpretation of
  the data of Ref. \onlinecite{Weichselbaum_and_White_DMRG} could be consistent with a
  system with no spin gap in the thermodynamic limit.

\subsubsection{Interchain Correlations}
\label{subsec:Interchain_Correlations}
Our analysis of intrachain correlations for systems of size $4 \times 4$, $4
\times 6$, $4 \times 8$ produced a clear and consistent picture of
antiferromagnetically ordered chains ($\theta_{J}=0$, $q_J=\pi$ for all chain
    lengths) accented by incommensurate interchain correlations.  The situation
for interchain correlations is not so simple.

The open question in the $J' \ll  J$ region is whether the systems exhibits a
one or two-dimensional spin liquid
phase\cite{Becca_Sorella_ED,Weng_Weng_Bursill_ED,Hauke_ED_MSWT,Chung_Marston_LSW_HATM}
or a collinear antiferromagnetic order driven by next-nearest chain
antiferromagnetic correlations and order by disorder\cite{Balents_RG}.  This
debate can be better informed by a consideration of the interchain ordering
vector, $q_{J'}$ and the importance of next-nearest chain antiferromagnetic
interactions to the ground-state.

The twist $\theta_{J'}$ which minimizes the ground-state as a function of
system \emph{width} is found to be $\pi$ for $4 \times 4$, and $0$ for $6
\times 4$ and $8 \times 4$ for $J'<J'_c$.  This is clearly inconsistent with
any classical ordering vector $q_{J'}$.  This supports the belief that, for the
system sizes under consideration, any long-range classical incommensurate spiral order is
suppressed.  Previous studies which have shown the lack of long-range spiral order had a
potentially critical flaw in that they used periodic boundary conditions which
undoubtedly destabilize such orderings.  It is then interesting that a lack of
spiral order is still found when the system has complete freedom to adopt an
incommensurate ground-state.

It is important to remember that, although the long-range incommensurate ordering is
suppressed short-range incommensurate correlations are still present.  This is manifest by the
complete lack of any features of the minimum twist when calculated in the total $S^z=1$ sector around
the critical $J'_c$.  An implication of this is that the short-range behavior of correlation functions would
show the same incommensurate behavior above and below $J_c'$. It is then natural to consider how strong these
incommensurate interchain interactions are, and how they compare to the
predicted next-nearest chain antiferromagnetic interactions that would drive a
CAF ordering.

The strength of interchain correlations can typically be determined by examining
$\left\langle \hat{S}_{\mathbf{x},\mathbf{y}}
\hat{S}_{\mathbf{x},\mathbf{y}+y'} \right\rangle$.  However, the value of such
an analysis here is hindered by the small system sizes numerically available.
This deficiency turns out not to be so significant since the qualitative
information relating to the correlation between chains can be inferred from the
curvature of $\theta_{J'}$.  For completely decoupled chains the ground-state
energy will have no dependence on the interchain twist $\theta_{J'}$, similarly
if the minima in $\theta_{J'}$ that minimizes the ground-state energy is found
to be extremely shallow then it can be argued that the interchain correlations
are extremely weak.  Thus, by taking the second numerical derivative, in the
total-$S^z=0$ sector, we can construct a \emph{$J'-$twist susceptibility}:
\begin{equation}
\frac{\partial^2 E_{gs}}{(\partial \theta_{J'})^2}=\chi_{\theta_{J'}}.
\end{equation}
This susceptibility will
probe the strength of interchain correlations with a large value of $\chi_{J'}$
representing strong correlations and a small value of $\chi_{J'}$ representing
weak ones.

\begin{figure}[t]
\includegraphics[scale=0.5]{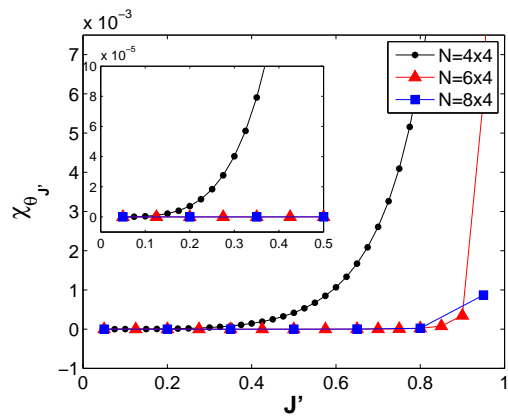}
\caption{\label{fig:chi_Tx_vs_Jp}  (Color online.) $\chi_{\theta_{J'}}$ vs. $J'$:
The curvature of the ground-state energy (i.e. total-$S^z=0$) about its minimum ($\theta_{J'}=\pi$ for $N=4 \times 4$, $\theta_{J'}=0$ for $N=6 \times 4$ and $N= 8\times 4$) with respect to $\theta_{J'}$
  ($\chi_{\theta_{J'}}$) versus $J'$ for systems of increasing width.  The size
  of $\chi_{\theta_{J'}}$ gives an indication of the strength and importance of
  interchain, (i.e. nearest chain) interactions to the ground-state energy.  It
  is clear that these correlations become smaller with width and become
  exceedingly weak as $J' \rightarrow 0$.  This is consistent with reasoning
  from RG and the lack of long-range spiral order in this region.} \end{figure}

The $\chi_{\theta_{J'}}$ dependence on $J'$ and system width can be seen in
Fig.~\ref{fig:chi_Tx_vs_Jp} for a $\delta \theta_{J'}$ of 0.1.  It can clearly
be seen that interchain correlations become tiny as the number of chains
increases.  In fact the interchain correlations are found to be zero within the
$10^{-13}$ precision of the numerics for systems of width 6 and 8 for small
$J'$ even for such a large value of $\delta \theta_{J'}$.  This is consistent
with the previous claim that these correlations are too weak to force spiral
ordering.  However, an RG analysis of the ATLHM\cite{Balents_RG,Kallin_Ghamari_RG}
posit that as the interchain correlations become weak with $J' \rightarrow 0$,
the next-nearest chains correlate antiferromagnetically with a strength,
$J_{nnc}$, which grows.  We will now consider the effect of such
correlations.

Recent series expansion work by Pardini and Singh in Ref.
\onlinecite{Pardini_and_Singh_Series_Expansion}  have suggested that an
incommensurately ordered ground-state has a lower energy than a CAF one for
small $J'$.  However, their work also showed that this energy difference was
extremely small and dependent on how short-ranged spiral correlations are
treated.  We found previously that the ground-state is not incommensurately
ordered for our system sizes for small $J'$ and instead exhibits intrachain
antiferromagnetism.  A relevant question is then whether the next-nearest chain
interactions are antiferromagnetic (CAF) or ferromagnetic (non-CAF or NCAF) and
whether these correlations grow as $J' \rightarrow 0$.  We previously determined
that $6 \times 4$ and $8 \times 4$ sized systems are minimized, in the total-$S^z=0$ subspace, by
$\theta_{J'}=0$. This observation makes it difficult to discriminate between CAF and NCAF phases
since both would have such a twist.  However, the $4 \times 4$ system is
minimized by a $\theta_{J'}=\pi$, which is inconsistent with CAF ordering.
This presents an opportunity to clearly demonstrate the effect of next-nearest
neighbour correlations.

\begin{figure}[t]
\includegraphics[scale=0.5]{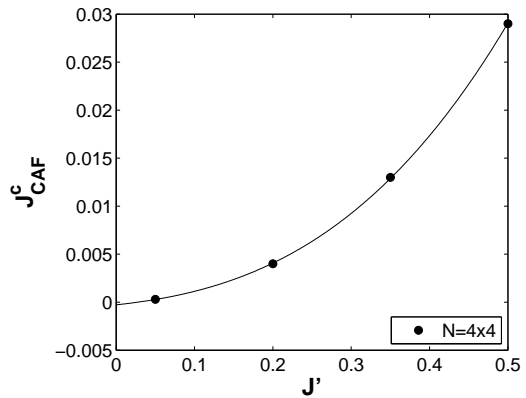}
\caption{\label{fig:J_CAF_vs_Jp}  (Color online.) $J_{CAF}^c$ vs. $J'$ for $N= 4 \times 4$:
As is discussed in the text, the $N= 4 \times 4$ system, whose $\theta_{J'}$ of
  $\pi$ is found to be incompatible with the predicted
  collinear-antiferromagnetic (CAF) ordering for $J' \ll 1$, can be forced to a
  $\theta_J$ of 0, consistent with this ordering by applying only a small
  next-nearest neighbour antiferromagnetic interaction $J_{CAF}$ (see text).
  Thus, as a demonstration of the subtle importance of these next-nearest chain
  interactions the critical $J_{CAF}^c$ for which $\theta_{J'}$ jumps from
  $\pi$ to $0$ is plotted as a function of $J'$.  For $J'=0.05$ this value
  becomes as low as $J_{CAF}^c=0.0003$ representing an extreme susceptibility
  to such interactions.  Conversely a next-nearest chain \emph{ferromagnetic}
interaction is found to have no effect on $\theta_{J'}$.  This suggests a
  strong preference for CAF order.  This solid line is given as an aid for the
  eye.} \end{figure}

We proceed by artificially inserting an exchange coupling between next-nearest
chains, $J_{nnn}\hat{S}_{\mathbf{x},\mathbf{y}}
\hat{S}_{\mathbf{x}-1,\mathbf{y}+2}$.  The question then is, at what strength
of $J_{nnn}$ does the $4 \times 4$ system adopt a $\theta_{J'}=0$ ordering
(which we take to be CAF).  This critical $J^c_{nnn}$ is shown, as a function
of $J'$ in Fig.~\ref{fig:J_CAF_vs_Jp}.  As $J' \rightarrow 0$ the necessary
``nudge'' the system needs to adopt a CAF ordering becomes very small. In fact,
for $J'=0.05$, this critical interaction strength is as tiny as 0.0003.
Contrarily, if a \emph{ferromagnetic} interaction is used (i.e. $J_{nnn} <0$)
then $\theta_{J'}$ does not change, regardless the magnitude of $J_{nnn}$.
The fact that such a minuscule increase in antiferromagnetic next-nearest
neighbour correlations can force the ground-state minimizing boundary twist
to jump to one consistent with CAF ordering and inconsistent with NCAF
ordering lends promise to the notion of CAF ordering in the thermodynamic limit.

The ability to easily force a $4 \times 4$ system into a CAF consistent
ordering is appealing, but hardly conclusive, evidence that CAF ordering will
occur for larger systems, especially since this system is so small.  We
therefore consider another means of analysing these interactions that can be
applied to larger systems.

\begin{figure}[t]
\includegraphics[scale=0.5]{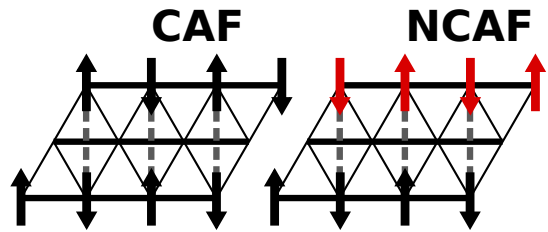}
\caption{\label{fig:CAF_Field_Lattice}  A diagram of the staggered fields
  applied in the generation of $\chi_{NNN}$, $\chi_{CAF}$ and $\chi_{NCAF}$.
    The field terms, $h \hat{S}^z_i$, are represented by arrows.  Fields are
    placed on every \emph{other} chain to allow the system freedom to adopt a
    classical $q_{J'}= \pi/2$ ordering.  The collinear antiferromagnetic (CAF)
    ordering is found on the left and corresponds to antiferromagnetic
    correlations between next-nearest chains.  For clarity, a sample
    next-nearest chain partner for the non-skewed triangular system is
    illustrated with a dotted line.  The non-collinear antiferromagnetic (NCAF)
    ordering, corresponding to \emph{ferromagnetic} next-nearest chain
    correlations, is shown on the right.} \end{figure}

We begin by perturbing our system with two different arrangements of staggered
magnetic field.  The first arrangement is chosen to be consistent with CAF
ordering and involves antiferromagnetic staggering along chains and between
next-nearest chains (see Fig.~\ref{fig:CAF_Field_Lattice}, left).  We only
apply fields to \emph{every other} chain in order to allow the system the
freedom to adopt the classical $q_{J'}=\pi/2$ ordering.  The second arrangement
is designed to be consistent with ferromagnetic ordering between next-nearest
chains (see Fig.~\ref{fig:CAF_Field_Lattice}, right) but still
antiferromagnetic along chains.  Two susceptibilities are constructed from this
perturbation.  

\begin{figure}[t]
\includegraphics[scale=0.5]{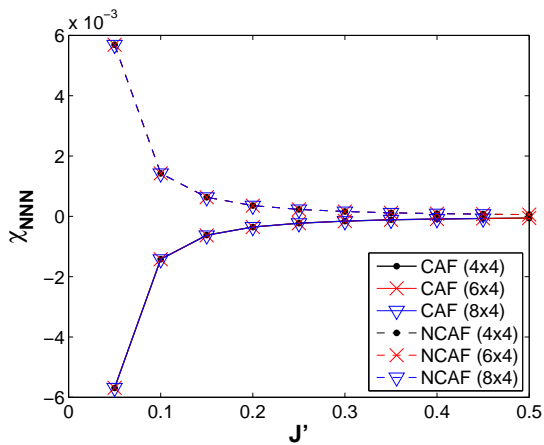}
\caption{\label{fig:NNN_Correlator_Susceptibility_vs_Jp}  (Color online.) $\chi_{NNN}$ vs. $J'$ and System Size:
$\chi_{NNN}$, being a susceptibility of the next-nearest chain correlation
  function to collinear (CAF) and non-collinear (NCAF) perturbative magnetic
  fields further discussed in the text, plotted against $J'$ for various system
  widths.  The perturbing field was of strength $h=0.001$ and applied to half
  the sites for both CAF and NCAF (see text).  Calculations were done in total-$S^z=0$ about the $\theta_{J'}=0$ minima.   This quantity is found to be
  largely system size independent and negative (positive) for CAF (NCAF)
  correlations.  This is consistent with the notion that a perturbative CAF
  field will cause the next-nearest chain correlations to grow more negative
  and a NCAF field to grow more positive.   The extremely similar magnitude of
  the two correlations suggest the system exhibits a delicate competition
  between collinear and non-collinear next-nearest chain correlations in the
  non-spiral ordered phase and that this susceptibility grows as $J'
  \rightarrow 0$.} \end{figure}

The first susceptibility, which we call $\chi_{NNN}$, is constructed from the
next-nearest neighbour chain correlation functions:
\begin{align}
\chi_{NNN} = \frac{\delta^2 \langle \hat{S}_{\mathbf{x},\mathbf{y}} \hat{S}_{\mathbf{x}-1,\mathbf{y}+2} \rangle }{\delta h^2}
\end{align}
where $h$ is arranged in one of the two (CAF or NCAF) ways.  The first
derivative term was found to be zero,  which could have been predicted on the
basis of spin inversion symmetry, and thus calculating this quantity is a
simple matter of numerical differentiation.  The results, as a function of
$J'$, are shown in Fig.~\ref{fig:NNN_Correlator_Susceptibility_vs_Jp} where
calculations were done only between chains that received a magnetic field (i.e.
    between chains 0 and 2 or 2 and 4 but not 1 and 3).   The specific chain
considered and the specific spin within that chain was found to be irrelevant.

This $\chi_{NNN}$ is found to be largely system size independent and negative
(positive) for CAF (NCAF) correlations.  This is consistent with the notion
that a perturbative CAF field will cause the next-nearest chain correlations to
grow more negative and a NCAF field to grow more positive.   The extremely
similar magnitude of the two correlations suggests the system exhibits a
delicate competition between collinear and non-collinear next-nearest chain
correlations in the non-spiral ordered phase and that this susceptibility grows
as $J' \rightarrow 0$.  The growth of this susceptibility, coupled with the
diminution of nearest-chain correlations as demonstrated in
Fig.~\ref{fig:chi_Tx_vs_Jp} seems consistent with the picture painted by
renormalization theory.  However, the system seems potentially equally
susceptible to ferromagnetic next-nearest chain interactions.  One then wonders
which of these correlations ultimately prevails.  In order to consider this we
consider yet another susceptibility.

To quantify the systems preference towards ferromagnetic versus
antiferromagnetic next-nearest chain ordering, we considered the effect that
perturbing magnetic fields of Fig.~\ref{fig:CAF_Field_Lattice} have on the ground-state energy.  We
thus define:
\begin{align}
\chi_{CAF} = \frac{\delta^2 E_{gs}}{\delta h_{CAF}^2}, \;  \; \: \chi_{NCAF} = \frac{\delta^2 E_{gs}}{\delta h_{NCAF}^2}.
\end{align}

\begin{figure}[t]
\includegraphics[scale=0.5]{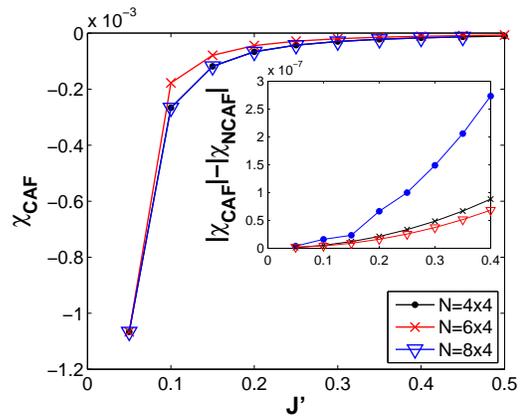}
\caption{\label{fig:Chi_E_CAF_vs_Jp_Plus_Inset}  (Color online.) $\chi_{CAF}$ vs. $J'$:
$\chi_{CAF}$, being the second derivative of the ground-state energy (i.e. total-$S^z=0$, $\theta_{J'}=0$) with
  respect to a collinear antiferromagnetic (CAF) perturbing magnetic field,
          versus $J'$ for systems of varying width.  For all system sizes the
            quantity is found to be negative and increasing in magnitude with
            decreasing $J'$.  This implies that the system's energy is lowered
            by promoting CAF-like ordering.  In order to establish the strength
            of this affinity for CAF order vs. non-collinear antiferromagnetic
            (NCAF) order, a similar ground-state susceptibility is defined
            relative to an NCAF perturbing magnetic field ($\chi_{NCAF}$).  The
            inset show the difference in magnitude between $\chi_{CAF}$ and
            $\chi_{NCAF}$.  $\chi_{CAF}$ is found to be greater for all system
            sizes and $J'$ though only by $\sim 10^{-7}$.} \end{figure}

As before, the first derivative term was found to be zero.  This is due to spin
inversion symmetry.  $\chi_{CAF}$ can be found plotted in
Fig.~\ref{fig:Chi_E_CAF_vs_Jp_Plus_Inset}. $\chi_{NCAF}$, which is not plotted,
  behaves identically except being positive.  The fact that $\chi_{CAF}$ is
  negative implies that CAF-ordering lowers the systems energy where NCAF,
  having a positive $\chi_{NCAF}$, increases it.  Furthermore, when we compare
  the \emph{magnitudes} of the two susceptibilities, which can be found in the
  inset of Fig.~\ref{fig:Chi_E_CAF_vs_Jp_Plus_Inset}, we see that $\chi_{CAF}$
  is in fact larger than $\chi_{NCAF}$ for all system sizes.  Though it is
  important to note that it is only larger by a margin of $\sim 10^{-7}$, and
  decreases with system width, further evidencing the tenuousness of these
  competing correlations. 

\section{Conclusion and Summary}
\label{sec:Conclusion_and_Summary}

In this paper we have demonstrated the power of twist boundary conditions to mitigate potentially disastrous finite-size effects in incommensurate systems.  Using these twisted boundary conditions we were able to extract the intrachain incommensurate $q$-vector $q_{J}$ and found it to be in good agreement with results on substantially larger systems.  Furthermore, we were also able to extract the \emph{inter}chain incommensurate $q$-vector $q_{J'}$.  To our knowledge our is the first work to allow fully incommensurate behaviour in both intra- and interchain directions.

Analysis of the incommensurability in both the ground-state and total $S^z=1$
excited state revealed a potential phase transition between a long-range spiral
ordered phase and one with short-range spiral correlations.  A scaling analysis
of this critical $J'_c$ suggests that this point is at $J'_c \sim 0.475$ for
systems of width 4 and $\sim 0.948$ for systems of infinite width (length 4).
We then attempted to characterize the $J'<J'_c$ phase.  We believe it to be
gapless in the thermodynamic limit, as well as dominated by both next-nearest
ferromagnetic and antiferromagnetic correlations.  Additionally, the nearest
chain correlations are found to become minuscule.  Further analysis reveals
that the antiferromagnetic interactions are marginally stronger in the systems
considered.  This is consistent with the renormalization group claim that this
region should be CAF ordered.

\begin{acknowledgments}

We would like to thank 
Sedigh Ghamari, Sung-Sik Lee and Catherine Kallin for many fruitful discussions.
We also acknowledge computing time at the Shared Hierarchical Academic
Research Computing Network (SHARCNET:www.sharcnet.ca) and research
support from NSERC.
\end{acknowledgments}

\bibliography{References}

\end{document}